# Zipper Model for the Melting of Thin Films

Mikrajuddin Abdullah[1,a], Shafira Khairunnisa[2], and Fathan Akbar[3]

[1]Department of Physics, Bandung Institute of Technology

Jalan Ganesa 10, Bandung 40132, Indonesia

[2]Department of Food Technology, Padjadjaran University,

Jalan Bandung–Jatinangor, Sumedang, Indonesia

[3]SMA Negeri 24 Bandung,

Jalan A.H. Nasution, Bandung, Indonesia

[a]Email: din@fi.itb.ac.id

## Abstract

We propose an alternative model to Lindemann's criterion for melting that explains the melting of thin films on the basis of a molecular zipper-like mechanism. Using this model, a unique criterion for melting is obtained. We compared the results of the proposed model with experimental data of melting points and heat of fusion for many materials and obtained interesting results.

## I. Introduction

The mechanism of material melting on a nanometer scale, such as the melting of nanoparticles, nanowires, and thin films, remains a topic of discussion. The most interesting phenomenon is the dependence of the melting temperature on the material dimension. The melting temperature decreases as the material dimension (diameter for nanoparticles and nanowires, or thickness for thin films) decreases. A reduction in melting temperature is observed when the dimension is below a critical size [1]. The critical sizes of materials are mostly below several tens of nanometers.

Many theories have been proposed to explain this phenomenon [1–8], most of which are based on Lindemann's criterion of melting [4–6]. Lindemann's criterion states that melting occurs when the root mean displacement of atoms in the crystal is a fraction of the atomic distance. Lindemann's criterion, however, does not give the exact fraction. We can demonstrate that a deviation in the fraction by several percent also leads to deviation in the predicted melting temperature by several percent. It is therefore important to propose a unique criterion for melting such that the predicted melting temperature is unique.

In this work we propose a different criterion for melting that is likely unique. The criterion is applied to explain melting of several thin films.

## II. Method

We apply the zipper model to explain the melting of a thin film deposited on a substrate. The zipper model has been discussed by Kittel to describe the transition of molecules [9]. At present, the thin film is modeled as a stack of ($n + 1$) identical sheets, each of which contains $N$ atoms. The melting process is assumed to start from the top sheet (free sheet) and develop towards the bottom sheet (the sheet in contact with the substrate) as illustrated in Fig. 1(a). We assign the top sheet the index $s = 0$ and the last sheet the index $s = n$. The melting process thus starts from the 0-th sheet. The $n$-th sheet never melts and is always in strong contact with the substrate. The $s$-th sheet melts only if all sheets from the 0-th sheet to the ($s − 1$)-th sheet have melted. If the $s$-th sheet has not yet melted then none of the sheets from the ($s + 1$)-th to the $n$-th sheet can melt. This is analogous to a zipper mechanism; that is, if a certain cell in the zipper is still closed then none of the subsequent cells can be opened as illustrated in Fig. 1(b).

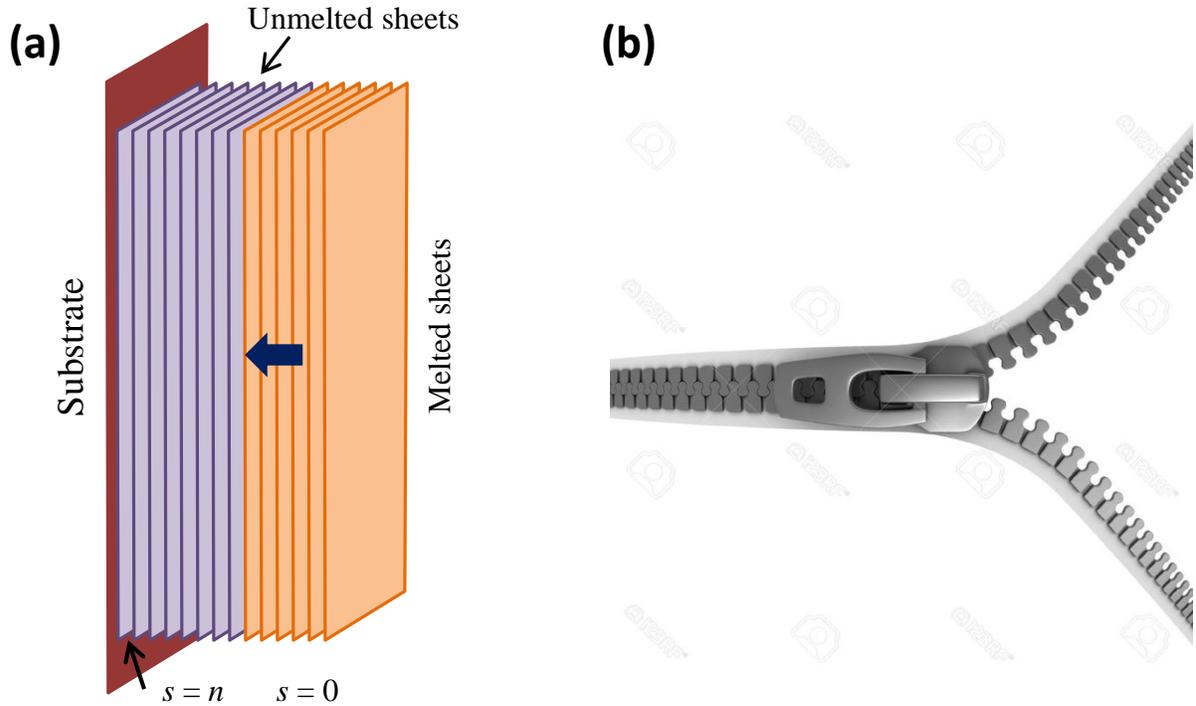

**Figure 1** (a) Development of the melting of a thin film and (b) illustration of a partially opened zipper (source of figure: www.123rf.com).

Suppose the energy required to change a sheet from the unmelted state to the melted state (the fusion energy of a sheet) is $E$ and the degeneracy of a sheet is $G$. We can express the partition function of the thin film similarly to the partition function of a molecular zipper as discussed by Kittel [9] as follows,

$$Z = \sum_{s=0}^{n-1} G^s e^{-sE/kT}. \tag{1}$$

Equation (1) is a simple geometric series and the result of the summation is

$$Z = \frac{1-x^n}{1-x}, \tag{2}$$

with

$$x = Ge^{-E/kT}. \tag{3}$$

We define here that melting occurs when the critical condition in Eq. (2) is achieved; i.e., the condition when $x = 1$. Therefore, if $T_m$ is the melting temperature, we have the relation

$$Ge^{-E/kT_m} = 1. \tag{4}$$

The fusion energy of a sheet can be written as $E = N\varepsilon$, where $\varepsilon$ is the fusion energy per particle.

Suppose the degeneracy of each atom in a sheet is $g$ and each atom is assumed to freely occupy its degeneracy state (with the states being uncorrelated with one another). This condition implies that

$$G = g^N. \tag{5}$$

According to Eqs. (4) and (5), we express the melting point of the thin film as

$$T_m = \frac{\varepsilon}{k \ln g}. \tag{6}$$

The fusion energy as shown in Eq. (6) is the fusion energy of an atom belonging to a thin film having thickness $d$. The fusion energy of an atom in a thin film should decrease as the film thickness decreases owing to the effect of the surface that completes the contribution of the volume. The binding energy of atoms located at the surface is lower than the binding energy of atoms inside the bulk. An increase in the fraction of atoms on the film surface reduces the average binding energy, implying a reduction in fusion energy. We propose an expression for the fusion energy of an atom in a thin film as

$$\varepsilon(d) = \frac{\varepsilon_\infty V - \alpha \sigma_s S}{V}, \tag{7}$$

where $\varepsilon_\infty$ is the fusion energy of an atom in the bulk material, $V$ is the volume of the thin film, $S$ is the surface area of the thin film, $\sigma_s$ is the surface energy of the film and $\alpha$ is a constant. Equation (7) might be speculative; however, it satisfies the condition that the effect of the surface vanishes for bulk materials. As $V = Sd$, we can write

$$\varepsilon(d) = \varepsilon_\infty \left(1 - \frac{\gamma}{d}\right), \tag{8}$$

where γ is a new constant. Substituting Eq. (8) into (6), we express the melting point of thin films as

$$T_m(d) = T_m(\infty)\left(1 - \frac{\gamma}{d}\right), \tag{9}$$

where

$$T_m(\infty) = \frac{\varepsilon_\infty}{k \ln g} \tag{10}$$

is the melting point of the material in its bulk state.

## III. Calculations and Confirmation

To confirm the acceptability of Eq. (10), we calculate the g-value for some materials. From data of fusion energies and melting points of metals we can calculate $g$. Equation (10) can be rewritten as $g = \exp(\varepsilon_\infty / kT_m)$. Table 1 lists the results of calculations for different metals: transition metals, alkaline metals, group IIIA–VIA elements, and lanthanide elements.

**Table 1**. Melting points [10–12] and calculated $g$ for transition metals, alkaline metals, group IIIA–VIA elements, and lanthanide elements

| Metal | Melting point (K) | Fusion energy/mol (kJ/mol) | Fusion energy/atom ($10^{-20}$ J/atom) | $g$ |
|---|---|---|---|---|
| Transition metals | | | | |
| Yttrium | 1799 | 11.4 | 1.89 | 2.14 |
| Iron | 1811 | 13 | 2.29 | 2.50 |
| Manganese | 1517 | 13.2 | 2.19 | 2.85 |
| Scandium | 1814 | 16 | 2.66 | 2.89 |
| Osmium | 3300 | 29.3 | 4.87 | 2.91 |
| Palladium | 1828 | 16.7 | 2.77 | 3.00 |
| Cobalt | 1768 | 16.2 | 2.69 | 3.01 |
| Silver | 1234 | 11.3 | 1.88 | 3.01 |
| Ruthenium | 2607 | 24.06 | 4.00 | 3.04 |
| Gold | 1336 | 12.5 | 2.08 | 3.08 |
| Tungsten | 3695 | 35.3 | 5.86 | 3.15 |
| Iridium | 2723 | 26 | 4.32 | 3.16 |
| Chromium | 2133 | 20.5 | 3.41 | 3.18 |
| Titanium | 1943 | 18.7 | 3.11 | 3.19 |
| Copper | 1357 | 13.1 | 2.18 | 3.20 |
| Niobium | 2743 | 26.8 | 4.45 | 3.24 |
| Platinum | 2043 | 20 | 3.32 | 3.25 |
| Mercury | 234 | 2.29 | 0.38 | 3.25 |
| Zirconium | 2127 | 21 | 3.49 | 3.28 |
| Nickel | 1726 | 17.2 | 2.86 | 3.32 |
| Vanadium | 2173 | 22.8 | 3.78 | 3.54 |
| Cadmium | 594 | 6.3 | 1.05 | 3.58 |
| Zinc | 692 | 7.35 | 1.22 | 3.59 |
| Tantalum | 3290 | 36.57 | 6.07 | 3.81 |
| Molibdenium | 2893 | 36 | 5.98 | 4.47 |
| Alkaline metals (IA and IIA groups) | | | | |
| Potassium | 336.3 | 2.33 | 0.39 | 2.30 |
| Cesium | 301.5 | 2.09 | 0.35 | 2.30 |
| Sodium | 370.83 | 2.6 | 0.43 | 2.32 |
| Rubidium | 312.5 | 2.19 | 0.36 | 2.32 |
| Rubidium | 312.5 | 2.19 | 0.36 | 2.32 |
| Calcium | 1115 | 8.56 | 1.42 | 2.52 |
| Barium | 1000 | 8 | 1.33 | 2.62 |
| Strontium | 1041.8 | 9.16 | 1.52 | 2.88 |
| Magnesium | 923 | 8.7 | 1.45 | 3.11 |
| IIIA – VIA groups | | | | |
| Thallium | 303.8 | 4.2 | 0.70 | 2.40 |
| Indium | 429 | 3.26 | 0.54 | 2.50 |
| Lead | 1755 | 13.8 | 2.29 | 2.58 |
| Selenium | 490 | 5.4 | 0.9 | 3.77 |
| Aluminum | 933 | 10.7 | 1.78 | 3.98 |
| Tin | 505 | 7 | 1.62 | 5.30 |
| Gallium | 303 | 5.59 | 0.93 | 9.21 |
| Bismuth | 544.4 | 10.9 | 1.81 | 11.13 |

| Boron | 2349 | 50 | 8.31 | 12.96 |
|---|---|---|---|---|
| Antimony | 630.8 | 19.7 | 3.27 | 13.79 |
| Tellurium | 449.5 | 17.5 | 2.91 | 18.46 |
| Polonium | 527 | 13 | 2.16 | 19.48 |
| Germanium | 1256 | 31.8 | 5.28 | 21.07 |
| Arsenic | 1090 | 27.7 | 4.60 | 21.32 |
| Silicon | 1714 | 50.2 | 8.34 | 33.97 |
| Lanthanide | | | | |
| Cerium | 1071 | 5.5 | 0.91 | 1.86 |
| Lanthanum | 1193 | 6.2 | 1.03 | 1.87 |
| Neodymium | 1297 | 7.1 | 1.18 | 1.93 |
| Praseodymium | 1204 | 6.9 | 1.15 | 1.99 |
| Gadolinium | 1585 | 10 | 1.66 | 2.14 |
| Samarium | 1345 | 8.6 | 1.43 | 2.16 |
| Dysprosium | 1685 | 11.1 | 1.84 | 2.21 |
| Terbium | 1629 | 10.8 | 1.79 | 2.22 |
| Ytterbium | 1092 | 7.7 | 1.28 | 2.35 |
| Europium | 1095 | 9.2 | 1.53 | 2.75 |
| Thulium | 1818 | 16.8 | 2.79 | 3.04 |
| Holmium | 1747 | 17 | 2.82 | 3.23 |
| Erbium | 1770 | 19.9 | 3.31 | 3.87 |

Interestingly, for the transition metals, the $g$-values in Table 1 are a nearly constant 3. As g is the degeneracy of an atom, this value probably relates to the three-dimensional vibration of the atom in the crystal. For alkaline metals, the $g$-value of alkaline metals is slightly smaller but nearly constant at around 2.5. Group IIIA–VIA elements have a larger $g$-value, mostly larger than 10.

A group of elements should have nearly the same degeneracy. As clearly shown in Table 1, the transition metal elements have nearly the same value of degeneracy. The alkaline metal elements also have nearly the same degeneracy. The lanthanide group has degeneracy values mostly between 2 and 4. Different behavior is observed in groups IIIA–VIA, for which there is a wide range of $g$-values. However, for these groups, the $g$-values are generally very large, mostly above 10.

We confirm the validity of Eq. (9) using experimental data. Figure 2 compares experimental data of the melting point of copper on a substrate of amorphous alloy Ta-W-N [13]. The bulk melting point of Cu is 1357 K. The curve in Fig. 2 was obtained from Eq. (9) using $\gamma$ = 4 nm. The comparison is fair, although some discrepancies are seen.

Figure 3 compares experimental data of the melting point of indium on a substrate of germanium [14]. The curve in Fig. 3 was obtained from Eq. (9) using $\gamma$ = 0.5 nm. The curve matches the experimental data well.

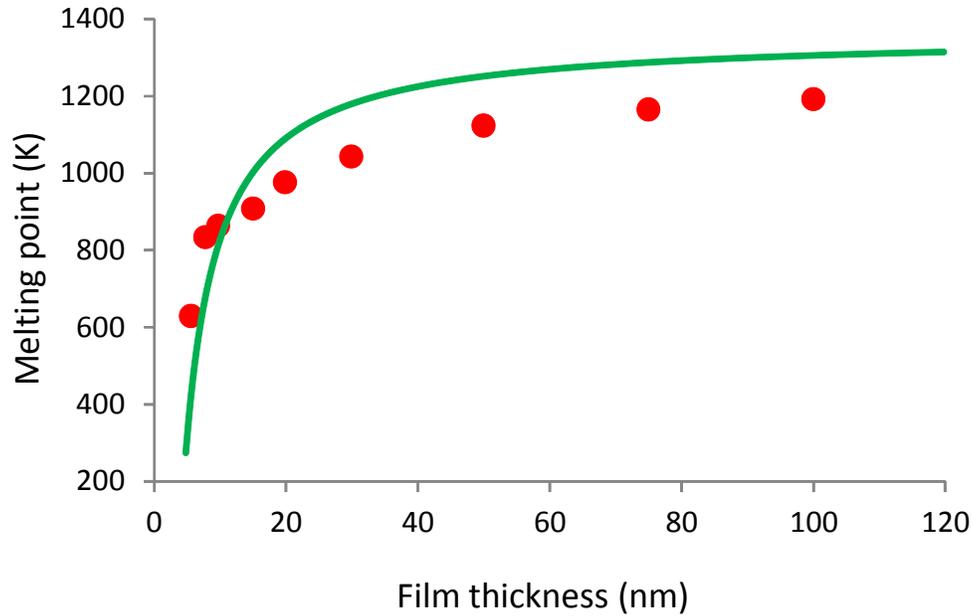

**Figure 2**. Melting temperature of copper film on a substrate of amorphous alloy Ta-W-N. Symbols are experimental data from [12], while the line is the curve calculated using Eq. (9) with $\gamma$ = 4 nm.

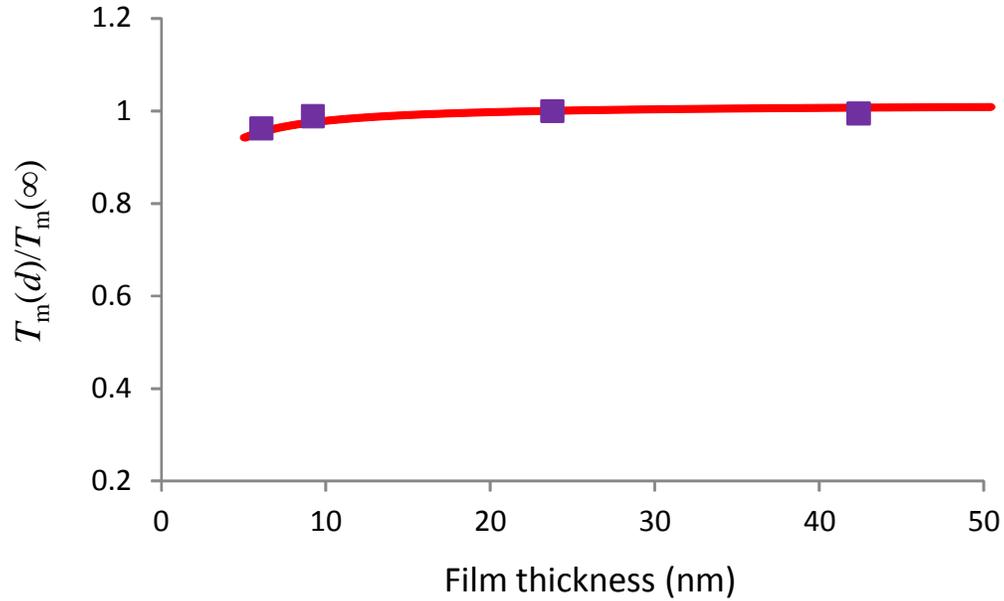

**Figure 3**. Melting temperature of indium film on a substrate of germanium. Symbols are experimental data from [13] while the line is the curve calculated using Eq. (9) with $\gamma = 0.5$ nm.

## Conclusion

The model of a molecular zipper proposed here was used to describe the melting of thin films. The model is able to explain the relationship between the heat of fusion and melting temperature and showed good consistency with experimental results for many metals. The degeneracies of atoms in a group of elements are nearly constant. Our simple model of the thickness dependence of the melting temperature of thin films well explains experimental data for copper and indium on substrates.

## Acknowledgements

This work was supported by a Pendaftaran Beasiswa Pendidikan Pascasarjana Dalam Negeri research grant (No. 310y/I1.C01/PL/2015) from the Ministry of Research and Higher Education, Republic of Indonesia, 2015.